\documentclass[onecolumn]{pasj00}
\draft

\begin{document}
\SetRunningHead{T.Okuda et al.}{Radiative Shocks in Rotating Accretion 
 Flows around  Black Holes}

\title{Radiative Shocks in  Rotating Accretion Flows 
around  Black Holes}

\author{Toru \textsc{Okuda}}
\affil{Hakodate College, Hokkaido University of
Education, 1-2 Hachiman-cho, Hakodate 040-8567}
\email{okuda@cc.hokkyodai.ac.jp}
\and
\author{ V. \textsc{Teresi}, E. \textsc{Toscano}, and 
 D. \textsc{Molteni}}
\affil{Dipartimento di Fisica e Tecnologie, Universita di Palermo,
 Viale delle Scienza, Palermo, 90128, Italy}
\email{vteresi@unipa.it}


\maketitle

\begin{abstract}
 It is well known that the rotating inviscid accretion flows with adequate 
 injection parameters around  black holes could form shock waves close 
 to the black holes, after the flow passes through the outer sonic 
 point and  can be virtually stopped by the centrifugal force.
 We examine numerically such shock waves in 1D and 2D accretion flows, 
 taking account of cooling and heating of the gas and radiation transport. 
 The numerical results show that  the shock location shifts outward
 compared with that in the adiabatic solutions and that the more rarefied 
 ambient density leads to the more outward shock location.
 In the 2D-flow, we find  an intermediate  frequency QPO 
 behavior of the shock location as is observed in the black hole candidate 
 GRS 1915+105.

\end{abstract}

\section{Introduction}
 Rotating inviscid and  adiabatic accretion flow around a black hole under a 
 pseudo-Newtonian  potentials can have two saddle type sonic points.
 After the inviscid flow with adquate injection parameters passes through the outer sonic points, 
 the supersonic flow can be virtually stopped by the centrifugal force, 
 forming a standing shock close to
 the black hole and again falling into the black hole supersonically.
@Apart from the initial works in the spherical transonic problems of accretion and wind,
 it was \citet{Fukue1987} under a full relativistic treatment 
 and Chakrabarti and his collaborators (\cite{Chakrabarti1989} ,
\cite{Abramowicz1990}, \cite{Chakrabarti1993}) under a pseudo-Newtonian 
potential who were the first to provide the satisfactory analytical or 
 numerical global shock solutions for transonic, invicid,
 rotating accretion around a Schwartzschild black hole.
 It has been shown that these generalized accretion flows could be responsible
 for the hard and soft state transitions or the quasi-periodic oscillation 
 (QPO) of the hard X-rays from the black hole candidates 
 (\cite{Molteni1et1996}, \cite{Ryu1997}, \cite{Lanzaf1998}).

 Recently, \citet{Das2001} examined analytically locations of sonic points and standing shocks
 in a thin, axisymmetric, adiabatic flow around a  black hole and 
  \citet{Das2002} and \citet{Das2003} showed that, 
 using four available pseudo-Newtonian potentials, the standing shocks 
 are essential ingredients in the multi-transonic black hole accretion disks. 
 Using the generalized multi-transonic accretion model,
\citet{Das1et2003} and \citet{Das2et2003} calculated analytically the QPO frequency of galactic
black hole candidates in terms of dynamical flow variables and  proposed a non-self-similar model
of coupled accretion-outflow system in connection to QPO of the black hole powered galactic
microquasars. 
 Most of these analytical or numerical works of shocks in the rotating accretion flows 
 around the black holes have been examined under the adiabatic condition,
 that is, the cooling and heating of gas and the radiation transport have not been taken account of.
 However, the radiation actually will play
 an important role in  the overall flow structure, especially the shock 
 location and the shocked temperatures.  Accordingly it is worth-while to examine the shocks 
 in the transonic black hole accretion disks, 
 taking account of the radiation effects.

 \section{ Model Equations}
 We examine  these shock problems firstly in one-dimension and secondly
 in axisymmetric two-dimensional inviscid flow. 
 A set of relevant equations consists of six partial differential 
 equations for density, momentum, and thermal and radiation energy.
 These equations include the heating and cooling
 of gas and radiation transport.   
 The radiation transport is treated in the gray, flux-limited diffusion
 approximation (\cite{Lever1981}).
 Using cylindrical coordinates ($r$,$z$,$\varphi$),
 the basic equations for mass, momentum, gas energy, and radiation energy 
 are written  in the following conservative form:

 \begin{equation}
   { \partial\rho\over\partial t} + {\rm div}(\rho\mbox{\boldmath$v$}) =  0,  
 \end{equation}
 \begin{equation}   
  {\partial(\rho v)\over \partial t} +{\rm div}(\rho v \mbox{\boldmath$v$})  = 
   \rho \left[ {{v_\varphi^2 } \over r}
  -{GM_* \over {(\sqrt{r^2+z^2}-r_{\rm g})^2 }}
   {r \over \sqrt{r^2+z^2}} \right] 
   -{\partial p\over \partial r}+f_r ,  
 \end{equation}
 \begin{equation}  
  {{\partial(\rho w)}\over \partial t} +{\rm div}(\rho w\mbox{\boldmath$v$}) 
 = -{\rho GM_* \over {(\sqrt{r^2+z^2}-r_{\rm g})^2}} {z \over \sqrt{r^2+z^2}}
  -{\partial p\over \partial z}
     + f_z , 
 \end{equation}
 \begin{equation}    
 {{\partial(\rho rv_\varphi)}\over \partial t} 
     +{\rm div}(\rho rv_\varphi\mbox{\boldmath$v$}) = 0 ,
\end{equation}
\begin{equation}  
  {{\partial \rho\varepsilon}\over \partial t}+
    {\rm div}(\rho\varepsilon\mbox{\boldmath$v$})
      = -p\;\rm div \mbox{\boldmath$v$} - \Lambda, 
\end{equation}
and 

\begin{equation}       
  {{\partial E_0}\over \partial t}+ {\rm div}\mbox{\boldmath$F_0$} +
        {\rm div}(\mbox{\boldmath$v$}E_0 +\mbox{\boldmath$v$}\cdot P_0) 
        = \Lambda 
      - \rho{(\kappa +\sigma)\over c}\mbox{\boldmath$v$}\cdot
      \mbox{\boldmath$F_0$} ,
 \end{equation} 
 where $\rho$ is the density, $\mbox{\boldmath$v$}=(v, w, v_\varphi)$ are the
 three velocity components, $G$ is the gravitational constant,
 $M_*$ is the central mass, $p$ is the gas pressure,
 $\varepsilon$ is the specific internal energy of the gas,  $E_0$ is 
 the radiation energy density per unit volume, and $P_0$ is the radiative
  stress tensor. It should be noticed that the subscript "0" denotes
  the value in the comoving frame and that the equations are correct
   to the first order of $\mbox{\boldmath$v$}/c$ (\cite{Kato1998}).
 We adopt a pseudo-Newtonian potential (\cite{Paczy1980})
 in equations (2) and (3), where $r_{\rm g}$ is the Schwartzschild radius
 given by $2GM_*/c^2$.
 The force density $\mbox{\boldmath$f$}_{\rm R}=(f_r,f_z)$ exerted 
 by the radiation field is given by
\begin{equation} 
  \mbox{\boldmath$f$}_{\rm R}=\rho\frac{\kappa+\sigma}{c}\mbox{\boldmath$F_0$}, 
\end{equation} 
 where $\kappa$ and $\sigma$ denote the absorption and scattering 
 coefficients and $\mbox{\boldmath$F_0$}$ is the radiative flux 
 in the comoving frame.
 For one-dimensional form, we put z=0 in equation (2) 
 and omit equation (3).
 
 The quantity $\Lambda$ describes the cooling and heating of the gas,
  i.e., the energy exchange between the radiation field and the gas
 due to absorption and emission processes,
 \begin{equation}      
      \Lambda = \rho c \kappa(S_*-E_0), 
\end{equation}
 where $S_*$ is the source function and $c$ is the speed of light. 
 For this source function, we assume local thermal equilibrium $S_*=aT^4$, 
 where $T$ is the gas temperature and $a$ is the radiation constant.
 For the equation of state, the gas pressure is given by the ideal gas law, 
 $p=R_{\rm G}\rho T/\mu$, where $\mu$ is the mean molecular weight 
 and $R_{\rm G}$ is the gas constant. 
  The temperature $T$ is proportional to the specific
 internal energy, $\varepsilon$, by the relation $p=(\gamma-1)\rho\varepsilon
  =R_{\rm G}\rho T/\mu$, where $\gamma$ is the specific heat ratio.  
  To close the system of 
 equations, we use the flux-limited diffusion approximation (\cite{Lever1981}) 
 for the radiative flux:
\begin{equation}
   \mbox{\boldmath$F_0$}= -{\lambda c\over \rho(\kappa+\sigma)}
   {\rm grad}\;E_0, 
\end{equation}
\noindent and
\begin{equation}
   P_0 = E_0 \cdot T_{\rm Edd}, 
\end{equation}
where  $\lambda$ and $T_{\rm Edd}$ are the {\it flux-limiter} and the 
 {\it Eddington Tensor}, respectively, for which we use the approximate
 formulas given in \citet{Kley1989}.
 The formulas fulfill the correct
 limiting conditions in the optically thick diffusion limit and the
 optically thin streaming limit, respectively.
 
 \begin{table}
\caption{Injection Parameters}\label{tab:table1}
\begin{center}
\begin{tabular}{lllllllll}
\hline\hline
 Case & $v_{\rm out}$ & $a_{\rm out}$&
  $\lambda_{\rm out}$ & $\epsilon_{\rm out}$& 
 $R_{\rm in}/R_{\rm g}$ & $R_{\rm out}/r_{\rm g}$
 & $\phi$  \\
 \hline
 1D  & 0.0751 & 0.0564& 1.875  &  0.0075 &
 2.0  & 100 & -- \\
 2D  & 0.0938 & 0.0738 &1.64 & 0.005 &  
 1.5  & 30 & $ 29^{\circ}$  \\
 \hline
 \end{tabular}
\end{center}
\end{table}

\section{ Model Parameters and Numerical Methods}
 For the central black hole, we assume a Schwartzschild black hole
   with mass  $M_*=10M_{\solar}$.
  We try to find the steady solutions with shocks by solving the time-dependent
  equations (1)--(6), which are numerically integrated by a finite-difference 
  method under initial and boundary conditions. To do this, we have to
  determine the injection parameters  such as the specific angular momentum
  $\lambda_{\rm out}$, the radial velocity $v_{\rm out}$, and the sound
  velocity $a_{\rm out}$ at an outer boundary radius $R_{\rm out}$, whose 
  parameters can lead to  a  shock wave close to the black hole.
  We search analytically these injection parameters through the examination 
  of the paremeter space ($\epsilon, \lambda$), where $\epsilon$ and $\lambda$ 
  are the total specific energy  and the specific angular momentum ( e.g. 
  \cite{Chakrabarti1989},\cite{Molteniet1994}, \cite{Molteniet1999}).
  Typical injection parameters for  1D and 2D flows obtained thus are 
  listed in table \ref{tab:table1}.
  Here, the velocities and distances are given in units of the speed of light
   $c$ and the Schwartzschild radius $r_{\rm g}$, respectively. 
  $\phi$ is the subtended angle of the centeral star  to the initial disk  at  
  $r=R_{\rm out}$, that is, tan $\phi$ =  $(h/r)_{\rm out}$, where $h$ is the
  disk thickness.
  2D case in the table corresponds to a super-Eddington 
  accretion  with an accretion rate $\dot M \sim 
  10^{20}$ g s$^{-1}$ which is $\sim 64 \dot M_{\rm E}$ for 
  the Eddington luminosity $L_{\rm E}= \dot M_{\rm E}c^2$, where $\dot M_{\rm E}$ 
  is the Eddington critical accretion rate, and 
  the ambient density $\rho_{\rm out}$ is taken to be $1.25 \times 
  10^{\rm -6}$ g cm$^{-3}$.

\begin{figure}
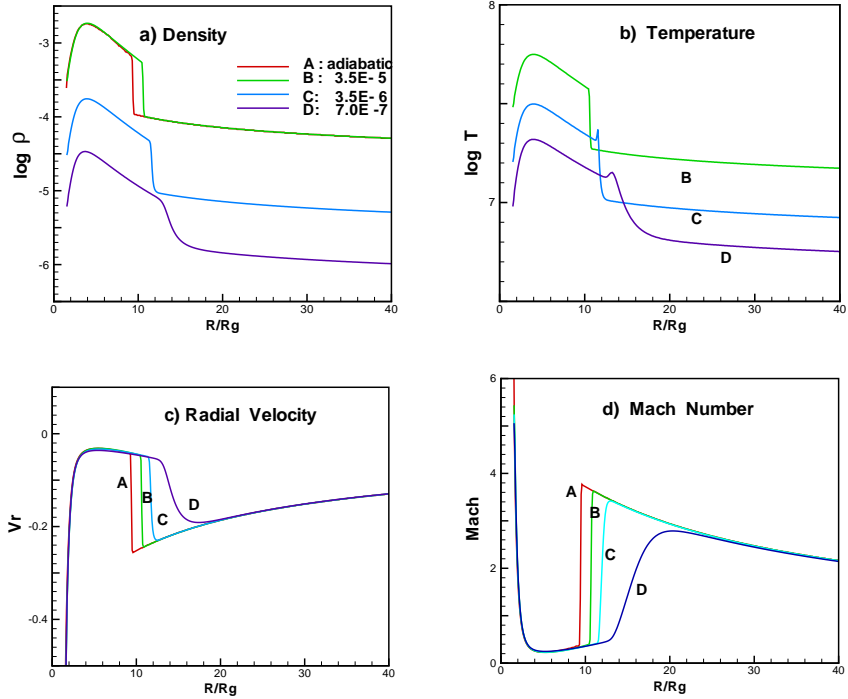

 \begin{center}
 \FigureFile(120mm,100mm){fig1.eps}
 \end{center}
  \caption {Flow structures in one dimensions with shock waves: 
  (a) density $\rho$: red line (A) shows the structure in adiabatic case
   with $\rho_{\rm out}=3.5\times 10^{-5}$ g cm$^{-3}$, and
   green (B), blue (C), and dark blue (D) lines denote the structures
   of non-adiabtic case with $\rho_{\rm out} = 3.5\times 10^{-5}, 3.5\times 
   10^{-6}$, and $7.0\times 10^{-7}$,
  (b) temperature $T$: case A is not included here, 
  because the temperatures in case A are very hot as $\sim 10^{10}$ --
    $10^{11}$ K,
  (c) radial velocity $v$, 
  (d) Mach number of the velocity  $v$.  
 }
 \label{fig:fig1}
 \end{figure}   

  At the outer boundary radius the injection parameters are always kept
   constantly. 
  At the inner boundary radius $R_{\rm in}$, a supersonic radial velocity 
  is specified.
  Adequate initial conditions of isothermal temperature and adequate 
  $r$-dependent distribution of density are imposed on the flow.
  With these initial and boundary conditions, we perform  time 
  integration of equations (1)--(6) until the steady solutions are obtained.
  The numerical schemes used are basically the same as that described 
  previously (\cite{Kley1989},\cite{Okuda1997}) but with no viscosity 
  description. 
  
\section{Numerical Results}
\subsection { One-Dimensional Flows}
  Firstly, under the injection  parameters in Table 1, we numerically 
  obtained a steady adiabatic shock with an ambient density $\rho_{\rm out}
  = 3.5\times 10^{-5}$ g cm$^{-3}$ at the outer boundary $r=R_{\rm out}$.
  If a set of the injection parameters are given, the shock position in the 
  adiabatic case is independent on the ambient density, because 
  the basic equations of the adiabatic and inviscid flow are almost 
  independent on the density as far as the ideal gas law is used.
  However, in the non-adiabatic cases, we obtain  steady 
  or sometimes nonsteady shock structure dependent on the ambient density
  and the shock locations are also dependent on the ambient density.  
  Figure \ref{fig:fig1} show the numerical results of density (a), 
  temperature (b), radial velocity (c), and Mach number (d) of the velocity 
  in the adiabatic flow (A) and non-adiabatic flows with different ambient 
  densities of $\rho_{\rm out} =3.5\times 10^{-5}$ (B), 
  $3.5\times 10^{-6}$ (C), $7.0\times 10^{-7}$ (D) g cm$^{-3}$, respectively. 
  The numerical results in the adiabatic flow agree 
   well with the analytical ones, where the shock locates at $r=9.4r_{\rm g}$. 
   
   The shocked temperatures in the adiabatic flow
   are too high to be $\sim 10^{10}$ -- $10^{11}$ K, while  actual
   postshock temperatures in B, C, and D are rather low as 
   2 -- 5 $\times 10^{7}$ K. The large temperature differences are 
   attributed to the differences of the temperatures specified initially at the outer 
   boundary, where same injection parameters of the sound 
   velocity $a_{\rm out}$ and the radial velocity $v_{\rm out}$  are used
   in the adiabatic and non-adiabatic cases.  If an optically thick
   and radiation-pressure dominant accretion flow is considered in the 
   non-adiabatic case, we have
   
   \begin{equation}
    (a_{\rm out}*c)^2 = \left[R_{\rm G} T_{\rm ad}/\mu \right]_{\rm adiabatic} 
      = \left[a T_{\rm nad}^4 / 
      {3 \rho_{\rm out}} \right]_{\rm non-adiabatic},
   \end{equation}
   where $T_{\rm ad}$ and $T_{\rm nad}$ are the temperatures at the outer
   boundary in the adiabatic and non-adiabatic cases, respectively.
  Therefore, for a given $a_{\rm out}$, we have smaller $T_{\rm nad}$ for
   smaller ambient density $\rho_{\rm out}$
  and, under the injection parameter $a_{\rm out}=0.0564$ in 1D-flow with 
  $\rho_{\rm out}=3.5\times 10^{-5}$
   g cm$^{-3}$, $T_{\rm ad} \sim 2\times 10^{10}$ K and 
   $T_{\rm nad} \sim 1.4\times 10^7$ K.  $T_{\rm ad}$ is three orders of 
   magnitude larger than $T_{\rm nad}$.
   Although the temperature profile in the non-adiabatic case 
   is dependent on the ambient density,
    the radial velocities and their Mach numbers take the radial profiles 
    independent on the ambient density in the pre-shock region, as is found in 
   Figure 1-(c) and (d). 
   
    Figure \ref{fig:fig1} also shows that the more rarefied ambient density 
    the flow has, the shock position moves more outward and the shock 
    thickness becomes broader. 
    The shock locations shift to 10.6, 12.0, and $\sim$ 15, in cases 
    B, C, and D, respectively.
    If the ambient density $\rho_{\rm out}$ is too low as 
    $\sim 3.5\times 10^{-7}$ g cm$^{-3}$, the steady shock does not 
    exist finally. 
   At the Rankine-Hugoniot relation of a standing shock, the pressure
  balance is supported by the dominant radiation pressure and 
  the ram pressure in the present non-adiabatic case.
  If the ambient density is taken to be smaller than that at a standing 
  shock, the pressure at the upstream just before the shock
   becomes much smaller because of the lower  temperature in the pre-shock
    region.
   Therefore, to set up a new pressure balance condition at the shock,
   the shock must shift outward  as far as the same injection parameters 
   are concerened. 
   When the shock position shifts more outward, the Mach number of the pre-shock velocity
   decreases, that is,  the shock becomes weaker. 
  Generally the shock thickness is inversely proportional to the wave strength,
  and the scale factor is the radiation mean free path in our case.
   Therefore the shock broadens gradually with decreasing ambient densities.
   If the ambient density is too low, it may be 
    imposibble to establish the Rankine-Hugoniot shock conditions in the 
   region considered and the shock disappears.

 \subsection{ Two-Dimensional Flows}
  Figures 2 and 3 show the temperature 
  contours for 2D adiabatic 
  and nonadiabatic accretion flows, respectively, under the injection
  parameters in Table 1 and the ambient density $\rho_{\rm out}= 1.25\times 
  10^{-6}$ g cm$^{-3}$ at the outer boundary.

  There are generally two important features of a two-dimensional rotating accretion
  flow around a black hole. One is the funnel wall, which is roughly 
  characterized by a surface ($x_{\rm f},z_{\rm f})$ of vanishing effective potential
   
   \begin{equation}
     \Phi_{\rm eff}={- 1 \over {(r_{\rm f}-1)}}+
           { \lambda^2 \over {{x_ {\rm f}}^2}}=0,
   \end{equation}
   where $r_{\rm f}=(x_{\rm f}^2+z_{\rm f}^2)^{1/2}$. The other is the centrifugal barrier 
   ($x_{\rm cf},z_{\rm cf}$), which is governed by the competition between the
   centrifugal force and gravitational force

 \begin{figure}
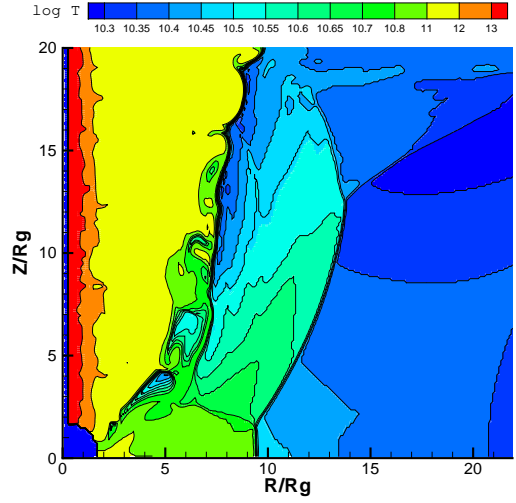

 \begin{center}
 \FigureFile(80mm,60mm){fig2.eps}
 \end{center}
 \caption {Temperature contours in two-dimensional adiabatic flow 
  with shock wave. The steady shock is formed at $r/r_{\rm g} \sim 9.3$ 
  near the equatorial plane and the shock front indicated by the thick black 
  lines extends obliquely to the upstream.  
 }
 \label{fig:fig2}
 \end{figure}
 
 \begin{figure}
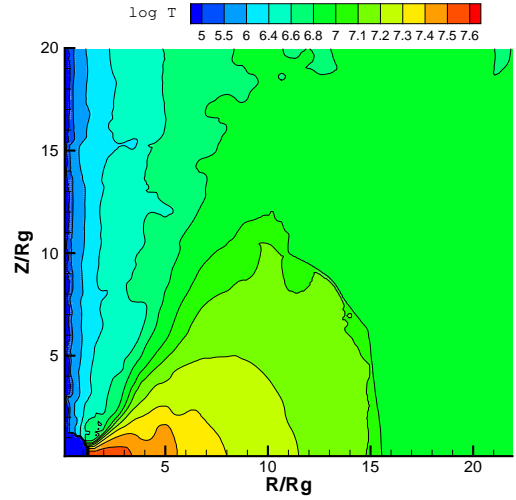

 \begin{center}
 \FigureFile(80mm,60mm){fig3.eps}
 \end{center}
   \caption {
  Same as figure \ref{fig:fig2}  but here the cooling and heating of the gas 
  and 
  the radiation transport are taken account of.  In this figure, the shock 
  appears at $r/r_{\rm g}  \sim 15 $ near the equatorial plane but 
  is not steady, and its position oscillates between 12 $\leq r/r_{\rm g} 
  \leq $ 17. 
 }
 \label{fig:fig3}
 \end{figure}

   \begin{equation}
      {-1 \over {(r_{\rm cf}-1)^2}} + {\lambda^2 \over x_{\rm cf}^3}=0.
   \end{equation}
  In the adiabatic flow, the standing shock appears at $r/r_{\rm g}=9.3$ 
 on the equatorial plane and is bent upward toward the upstream, roughly 
 following the contour of the centrifugal barrier. The post-shock temperatures
 are  as high as $\sim 10^{10}$ -- $10^{11}$ K as is found in 1D case.
  In the red and yellow regions between the rotating axis and the funnel wall, 
 the temperatures are much higher and the densities are very low and in these
  regions the outflow gas with high velocities are formed.
  These results are  similar to that by the previous TVD and SPH 
  simulations (\cite{Molteni2et1996}) in the 2D adiabatic flow.

\begin{figure}
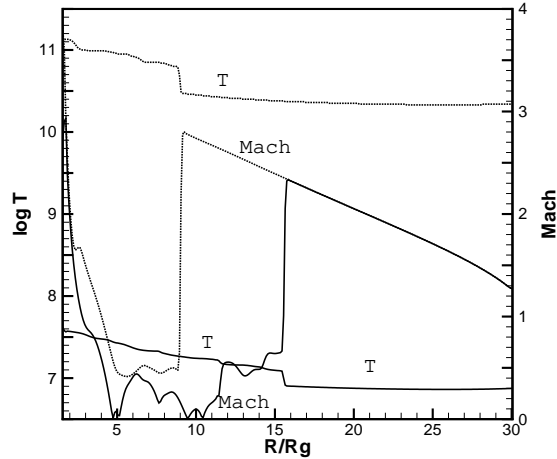

 \begin{center}
 \FigureFile(80mm,60mm){fig4.eps}
 \end{center}
 \caption {Profiles of the temperature and the Mach number of the radial 
 velocity on the equatorial plane for 2D adiabatic (dotted lines) and 
 nonadiabatic (solid lines) flows.
 }
 \label{fig:fig4}
 \end{figure}

 In contrast to the adiabatic flow, when the cooling and heating of gas and the
  radiation are taken account of, a spheroidal shock is formed between
  the funnel wall and the equatorial plane and the shock oscillates 
  quasi-periodically. 
  Figure \ref{fig:fig3} shows the temperature contours at the evolutionary 
  time $t = 5.5\times 10^4 r_{\rm g}/c$  in the non-adiabatic 
  flow where the shock appears at $r_{\rm s} \sim 15 r_{\rm g}$ on the 
  equatorial plane. The overall flow is optically thick except the cone-like 
  funnel region between the funnel wall and the rotational axis.
  In the funnel region we have very rarefied and optically thin flow 
  but the temperatures are not so high compared with the high temperatures
   in the viscous accretion flows where the viscously dissipative energy
   heats up considerably (\cite{Okuda2002}). 
   As a result, a relativistically high velocity jet is not formed here 
   in spite of the high input accretion rate. 
   The temperatures in the shocked region are
   1 -- 4 $\times 10^7$ which are three 
 orders of magnitude smaller than that in the adiabtic flow. 
  For clearness of the shock location and the shocked temperature, 
  in figure \ref{fig:fig4}  
 we plot the temeprature and  the mach number of the radial
 velocity on the  equatorial plane. It is clear that the effects of 
 radiation shifts  the shock position $r_{\rm s} \sim 9r_{\rm g}$ 
 to $\sim 15r_{\rm g}$ and the shock strength weakens.
 
\begin{figure}
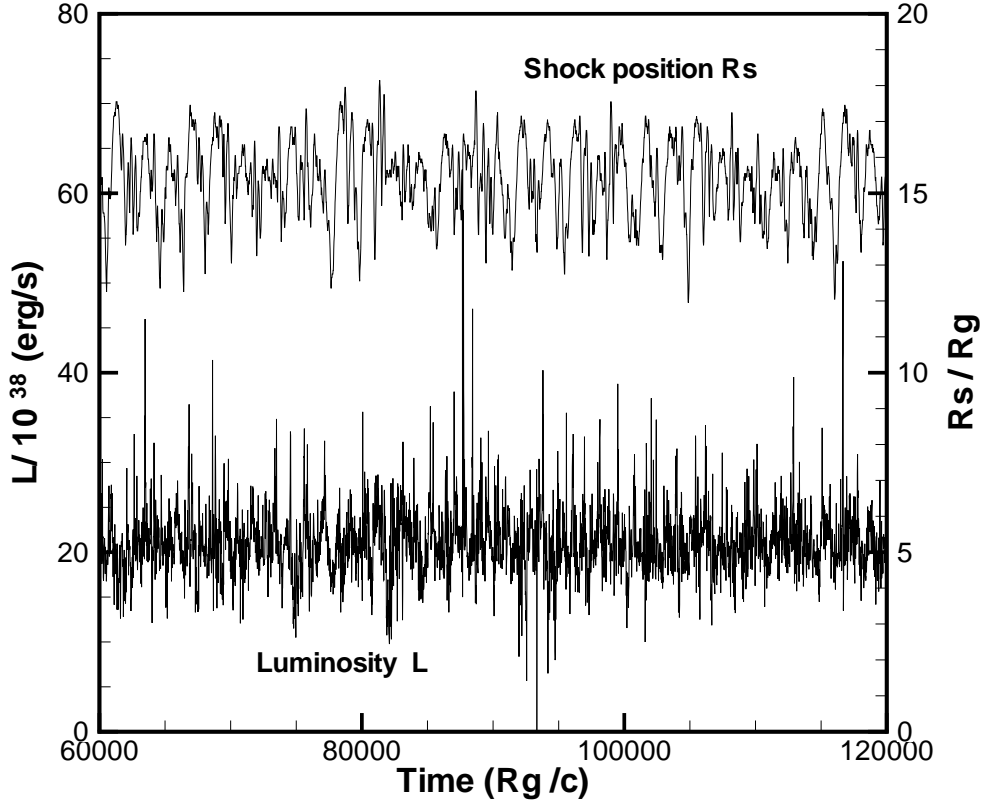

 \begin{center}
 \FigureFile(140mm,60mm){fig5.eps}
 \end{center}
 \caption {Shock location and  luminosity 
 versus as a function of time in units of $r_{\rm g}/c$  in 2D nonadiabatic 
  flow.
 }
 \label{fig:fig5}
 \end{figure}

 The shock in figure \ref{fig:fig3} oscillates  and this induces the 
 luminosity fluctuations.
 Figure \ref{fig:fig5} shows the shock radii on the equatorial plane 
 and the luminosity as a function of time in units of $r_{\rm g}/c$. 
 Although the time variation of the shock position on the equatorial
  plane is complicated, we find that it
 oscillates quasi-periodically with a  period of $\sim 2000 r_{\rm g}/c$ 
 (0.2 sec) and its frequency $\nu \sim$ 5Hz. The spheroidal shock also
 oscillates in the similar way  but the luminosity shows a more complicated
 variation due to the convective nature and the geometry of 2D-flow. 
 
\section{Discussion and Conclusion}
 In the present paper, we have numerically examined 1D and 2D inviscid 
 transonic flows, taking account of the cooling and heating of  gas and 
 radiation transport. As the results, we find that the location of the
 centrifugally driven shock  drifts more outward from the black
  hole  compared with that in the adiabatic flow and that the postshock 
  temepratures are as high as 1 -- 4 $\times 10^7$ K which is three orders 
  of magnitude lower than that in the adiabatic flow. When a set of the 
  injection parameters such as the specific angular momentum $\lambda$, 
  the radial velocity $v_{\rm out}$, and the sound velocity $a_{\rm out}$ 
  at the outer boundary are specified, the more rarefied ambient density the 
  flow has, the  shock location  drifts more outward.  When the ambient 
  density is too low, the shock wave  exists no longer under the injection
  parameters.  Depending on the injecteion parameters, even if the shock
  wave exists, it becomes sometimes unstable, that is, the shock location
  oscillates. 
  
  The transonic accretion shocks around the black holes have been applied
  to the hard and soft spectral states observed in  the black hole candidates 
  (\cite{Chakrabarti1995},\cite{Chakrabarti1997}). 
  In the shock model, the accretion disk is decomposed into three distinct
   components : (1) an optically thick
   Keplerian disk on the equatorial plane ($r > r_{\rm s}$); 
   (2) a sub-Keplerian optically thin halo above the disk ($r > r_{\rm s}$);
   (3) a hot, optically slim post-shock region ($r < r_{\rm s} \sim 5$ -- 
   $ 10 r_{\rm g}$). The hot, dense post-shock region intercepts soft photons 
   from a pre-shock region and a cold outer accretion disk and reprocesses
    them to form high-energy photons. 
   This model seems to explain well the observed properties. 
    In this respect, our results of the radiative shock may have such 
    observational appearance in black-hole candidates.
   \citet{Molteni1et1996} and \citet{Lanzaf1998} have examined shock waves in
  the viscous accretion disks around black holes and  showed that when the 
  viscosity parameter $\alpha$ is less than a critical value, 
  standing shock waves are formed but that if the viscosity is high then 
  the shock wave dissapears.
  Also in our 1D-cases, we confirmed that the standing shock is formed for 
  small $\alpha$  and  that the 
  shock position drifts more outward for  larger viscosity parameter
  until it attains to $ \sim 10^{-4}$.

 The shocked accretion flows have been applied to the QPO phenomena in
 galactic black hole candidates.
 The galactic X-ray transient source GRS 1915+105  exhibits 
 various types of quasi-periodic oscillations with frequencies ranging from
  $\sim$ 0.001 -- 10 Hz to $\sim$ 67 Hz (\cite{Morgan1997}).
  Analyzing X-ray data in  GRS 1915+105, \citet{Rao2000}
   showed that a thermal-Compton spectrum is responsible for the QPO
   generation and supported the hypothesis that the QPO is generated by the
   oscillation of the shock front in the transonic accretion flow. 
  The QPOs are classified as (1) the low-frequency QPO ($\nu_{\rm L} \sim $
   0.001 -- 0.01Hz), (2) the intermediate frequency QPO ($\nu_{\rm I} \sim$
    1 -- 10Hz), and (3) the high frequency QPO ($\nu_{\rm H} \sim$
     67 Hz).
  The QPO-like behavior of the shock
   oscillations with $\nu \sim $ 5 Hz  in our 2D-flow may  be 
  representative of the intermediate frequency QPO observed in GRS 1915+105.
 The  oscillation amplitude of the shock location could be up to 15 \% of 
 the distance of the shock wave from the black hole.
  This results in a quasi-periodic behavior of the luminosity with an
 amplitude variation of a factor 2. However
  we are not able to find  distinctly these frequency peaks of 
 $\nu_{\rm I}$, $\nu_{\rm L}$, and $\nu_{\rm H}$ from the power spectra 
 of the luminosity curve.
  
 \citet{Molteni1et1996}  have found the 
 QPO behaviors in viscous accretion disks
 with shock waves around a black hole with $M = 10^8 M_{\odot}$. 
 They obtained stable shock oscillations with periods of 1500 -- 2000 
 in units of $r_{\rm g}/c$ , which depends on an index of power-law 
 cooling function of gas. 
 The  oscillation period $t_{\rm osc}$ of 1500 units in the case of 
 bremsstrahlung cooling is comparable to 2000 units in our case.
 On the other hand, they interpreted the 
 oscillation period as the advection timescale $t_{\rm adv}$  only when
  it is comparable to the cooling timescale of the flow
  and estimate it as follows:
  
  \begin{equation}
    t_{\rm osc} \sim t_{\rm adv} =  \int_1^{r_{\rm s}}{ dr \over v_{\rm +}} ,
  \end{equation}
  where $v_{\rm +}$ is  the post-shock velocity.
  If the pre-shock velocity $v_{\rm -}$ at the shock is taken to be a fraction 
  $ f$ of the free-fall velocity, 
 $t_{\rm osc} = R r_{\rm s}^{3/2} /f$ where $R$ is the compression ratio
 of gas at the shock.  Using $r_{\rm s}/r_{\rm g}=17$, $R$= 7, and $f$ =0.7
 estimated in our results,  
  we have $t_{\rm osc} \sim$ 700 units which is  smaller by a few factor
  than actual 2000 units. 
  If a post-shock flow of constant velocity $ v_+ = v_{\rm 0}/R$ is used,
   where  $v_{\rm 0}$ is a fraction of the speed of light, $t_{\rm osc} 
   = R \;r_{\rm s} /v_{\rm 0}$.  For $r_{\rm s}=17, R= 7$, and $v_{\rm 0}=
   0.066$, we have $t_{\rm osc}$ = 1800 units 
   (0.17 sec) and a frequency of $\nu \sim $ 6 Hz  which corresponds to 
   the shock oscillation frequency in our 2D-flow.
 Assuming an outflow from the post-shock region ,
 \citet{Chakrabarti1999} derived a correlation between
 the shock location and the duration of the quiescence state
 of  GRS 1915+105 with typical QPOs and \citet{Chakrabarti2000} 
 showed that the correlation with $v_{\rm 0}=0.066$ was well fitted  
 to public-domain data from RXTE. 
 These transonic shock wave models in rotating accretion flows around black 
 holes may be promissing for the explanation of the QPOs in the black 
 hole candidates.Further examinations of the models  are required  
 from a 2D-simulation point of view.

\end{document}